\documentclass[twocolumn,letter]{jpsj2}

\def\Vec#1{\mbox{\boldmath $#1$}}

\title{
Boojums in Rotating Two-Component Bose-Einstein Condensates
}

\author{Hiromitsu Takeuchi\thanks{E-mail address:hiromitu@sci.osaka-cu.ac.jp} and Makoto Tsubota\thanks{E-mail address:tsubota@sci.osaka-cu.ac.jp}
}

\inst{%
Department of Physics, Osaka City University, Sumiyoshi-Ku, Osaka 558-8585, Japan
}

\recdate{\today}

\abst{%
A boojum is a topological defect that can form only on the surface of an ordered medium such as superfluid $^3$He and liquid crystals. We study theoretically boojums appearing between two phases with different vortex structures in two-component BECs where the intracomponent interaction is repulsive in one phase and attractive in the other. The detailed structure of the boojums is revealed by investigating its density distribution, effective superflow vorticity and pseudospin texture.
}

\kword{%
Bose-Einstein condensate, vortex, multicomponent, texture, interface, superfluid $^3$He}

\begin{document}
\maketitle

The study of topological objects is important because such objects appear not only in condensed matter physics but also in cosmology and high-energy physics \cite{Topolo}. Atomic-gas Bose-Einstein condensates (BECs) are ideal for examining topological defects and textures in a quantum condensed system \cite{KTUreview}. A major advantage of this system is that it can be accurately
described using the mean-field theory. Moreover, it is a good system to study experimentally because the atomic interaction is tunable through the Feshbach resonance and optical techniques allow one to control and visualize the condensates directly. Although topological defects and textures have been studied thoroughly in anisotropic superfluid $^3$He \cite{vollhardt,volovik}, interest in BECs with multicomponent order parameters has been increasing. For example, Kasamatsu {\it et al.}\cite{KTUreview} numerically simulated vortex structures in two-component BECs that were similar to those in superfluid $^3$He-A. Hence, we can learn about topological defects and textures in multicomponent BECs as well as in superfluid $^3$He. This sort of work can lead to remarkable developments toward the elucidation of topological objects in nature. In this study, we theoretically examine boojum structures in two-component BECs and show how they are analogous to boojums on the interface between A and B phases of superfluid $^3$He. This study pioneers the research field of interfacial topological defects of multicomponent BECs.

A boojum is a point defect that can exist only on the surface of an ordered medium. The name came from Mermin who analyzed these defects in superfluid $^3$He \cite{mermin}. Boojums at the interface separating A and B phases of superfluid $^3$He have been recently studied by Blaauwgeers {\it et al.} \cite{blaau} They made the A-B phase boundary in a cylindrical container by controlling the gradient of the magnetic field along the axis of the cylinder. When the container was rotated, each phase had a distinct structure of quantized vortices. In particular, the vortex core diameter of the A phase is 10$^3$-fold larger than that of the B phase. When the angular velocity $\Omega$ is smaller than a critical value $\Omega_c$, the A phase rotates by making doubly quantized vortices with "soft cores", whereas the B phase does not rotate. Also, the A phase vortices bend near the phase boundary toward the container wall \cite{hanninen}. In contrast, when $\Omega$ exceeds $\Omega_c$, the A phase vortices penetrate into the B phase, which can also rotate by forming single-quantized vortices with "hard cores". The vorticity distribution is continuous in the A phase, but singular at the cores of the vortices in the B phase. The boojums appear at the A-B phase boundary where the two kinds of vortices connect. However, we have little quantitative understanding of the structure of the order parameters around boojums. This lack of knowledge is partly a result of the difficulty in analyzing a system with such a large difference in coherence length between the two phases.

\begin{figure}[t] \centering
  \includegraphics[width=.95 \linewidth]{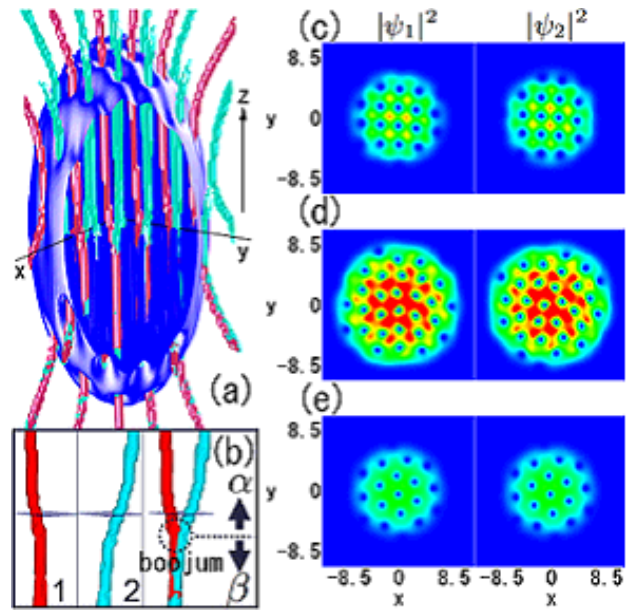}
  \caption{ Boojums in rotating two-component BECs.
 (a) The constant surface density $| \psi_1 | ^2 + | \psi_2 | ^2 $ is shown as the blue surface. The vortices of component 1 are red and those of component 2 are light green. (b) Vortices of component 1 (left) and component 2 (center), and superposition of two vortices(right). The gray planes are the $z=0$ planes. (c-e) Density distributions $| \psi_1 | ^2$ (left) and $| \psi_2 | ^2$ (right) in various cross sections perpendicular to $\hat{\Vec{z}}$. (c) Section at $z=10.8$. (d) Section at $z=-0.4$. (e) Section at $z=-11.2$. The box dimensions along the $x$, $y$, and $z$ axes are $-12.8$ to $+12.8$, $-12.8$ to $+12.8$, and $-25.6$ to $+25.6$, respectively ($256 \times 256 \times 512$ in discretized space). }
\end{figure}

Because of the similarity of the order parameters between superfluid $^3$He and a liquid crystal, boojums have also been studied in liquid crystals.\cite{kraus,park} Boojums are also reported in molecular patterns of fatty acids at an air-water interface \cite{riviere}. Then, atomic BEC, which is a well-controlled  quantum system,  should be a ground giving boojums. The research  in this direction may push out multicomponent BECs as a testing ground of the physics of pattern formation
 in which various topological defects play important roles \cite{Topolo}.

For this study, we numerically created two phases of different vortex structures in two-component BECs elongated along the rotation axis $\Vec{\Omega} = \Omega \hat{\Vec{z}}$, as shown in Fig. 1(a). The intercomponent interaction is tuned to be repulsive for $z>0$ and attractive for $z<0$. Thus, for repulsive interactions, the vortices of one component alternate with those of the other component, whereas the vortices of two components have common cores for attractive interactions \cite{vorsheet, JILA, Mueller_Ho}, as shown in Figs. 1(c) and 1(e). Then, just below the $z=0$ plane, bifurcation points of vortices form (Fig. 1(b)), which create boojums. The regions above and below the boojums are called the $\alpha$- and $\beta$-phases, respectively. By introducing an effective velocity and the pseudospin representation, we can more easily examine the structure of boojums and reveal their relationship to textures in superfluid $^3$He.

 We obtained stable states of two-component BECs in a frame rotating with an angular velocity $\Vec{\Omega}=\Omega \hat{\Vec{z}}$ by solving the time-independent, coupled Gross-Pitaevskii (GP) equations. For simplicity, we assume that $m=m_1=m_2$, $N=N_1=N_2$, and  $\omega_{\bot}=\omega_{1 \bot}=\omega_{2 \bot}$, where $m_i$, $N_i$ and $\omega_{i \bot}$ are the particle mass, the number of particles, and the radial trapping frequency of component $i$ ($i=1, 2$), respectively. It is convenient to measure the length, time, and energy scale in units of $b_{\bot}=\sqrt[]{\hbar / m \omega_{\bot}}$, $\omega_{\bot}$ and $\hbar \omega_{\bot}$, respectively. In these units, the coupled GP equations are 
\begin{eqnarray}
\mu_i \psi_i=\left( -\frac{1}{2}\nabla^2 +V -\Omega L_z+\Sigma_{j=1,2}
U_{ij} | \psi_j | ^2\right) \psi_i  \nonumber \\ 
\quad (i=1,2), \label{GPE}
\end{eqnarray}
 where $\mu_i $ is the chemical potential of component $i$, $L_z = -i(x \partial _y -y \partial_x)$ is the angular momentum operator, and the wave functions are normalized as $\int d\Vec{r} |\psi_i|^2=1$. The harmonic potential is $V=\frac{1}{2}(x^2+y^2+\lambda ^2 z^2) $ with anisotropic parameter $\lambda=0.3$. The intracomponent and intercomponent coupling constants respectively are $U_{11}=U_{22}=4\pi Na_{ii}/b_{\bot}$ and $U_{12}=U_{21}=4\pi Na_{12}/b_{\bot}$ with the corresponding $s$-wave scattering lengths $a_{ii}$ and $a_{12}$ being dependent on $z$.

In this system, it is essential for making boojums that BECs have the two regions with $U_{12}$ positive and negative.
Then we assume
 $U_{11} = U_{22} =1000 $ and $U_{12}=750 $ for $ z>0 $,
 $U_{11} = U_{22} =2500 $ and $U_{12}=-750 $ for $ z<0 $,
 and $\Omega=0.8$.
 For making stable boojum structures, it is neither necessary to change 
the interaction parameters so sharply nor to make $U_{11}$ and $U_{22}$ spatially dependent. Such a parameter setting is just for visualizing the structures of boojums clearly. The density distributions of each component are shown in Figs. 1(e)-1(j). The choices of $\Omega$ and $U_{12}$ were based on the phase diagram by Kasamatsu {\it et al.} \cite{vorsheet} such that the $\alpha$-phase vortices made the alternating square lattice that is shown in Fig. 1(c). In the $\beta$-phase, the two components have the same density distribution; that is, $ n_1 (\Vec{r} ) = n_2 (\Vec{r} ) $ where $ n_i (\Vec{r} ) = |\psi _i (\Vec{r} ) |^2 $. In this case, the two components are reduced to one component, thus making a triangular vortex lattice (see Fig. 1(e)) with a solid-body rotation except near the vortex cores with singularities. Thus, the vortices of each component are bifurcated on the interface between the two phases. The structure of one pair of two vortices is shown in Fig. 1(b). Here, a boojum exists at the point where the pair of two vortices is bifurcated.

The structures of the boojums of this system were revealed by investigating the
 distribution of the vorticity of the effective superflow velocity $\Vec{v}_{\rm eff}$. This velocity is defined as \cite{KTUeff}
 \begin{eqnarray}
 \Vec{v}_{\rm eff} = \frac{\Vec{j}_{1}+\Vec{j}_{2}}{n_1 +n_2},
 \end{eqnarray}
 where $\Vec{j}_i = n_i \nabla \theta_i $ and $ \theta_i $ is the phase of $ \psi_i $. From $\Vec{v}_{\rm eff}$, we obtain the effective vorticity $\Vec{\omega}_{\rm eff}$ as $ \Vec{\omega}_{\rm eff} = \nabla \times \Vec{v}_{\rm eff} $. The resulting three-dimensional structure of $|\Vec{\omega}_{\rm eff}|$ near the boojums is shown in Fig. 2(a), which may be called a boojum lattice. Figures 2(b)-2(d) show the distributions of the $z$ component $(\Vec{\omega}_{\rm eff})_z$ and the unit vector $\Vec{v}_{\rm eff}/|\Vec{v}_{\rm eff}|$ in three cross sections perpendicular to $\hat{\Vec{z}}$. In the $\beta$-phase, the vorticity distribution is similar to that of a one-component BEC with singularities at the vortex cores (Fig. 2(d)). We call such a singularity of $\Vec{\omega}_{\rm eff}$ "singular vortex". In contrast, in the $\alpha$-phase, the vorticity distribution is continuous, making a square lattice of fuzzy round blobs (Fig. 2(b)). Each blob has the vortex core of either component (red and blue points in Fig. 2(b)). We call such a blob "continuous vortex". A boojum appears at the point on the interface where a singular vortex splits into two continuous vortices.

\begin{figure}[t] \centering
  \includegraphics[width=.95 \linewidth]{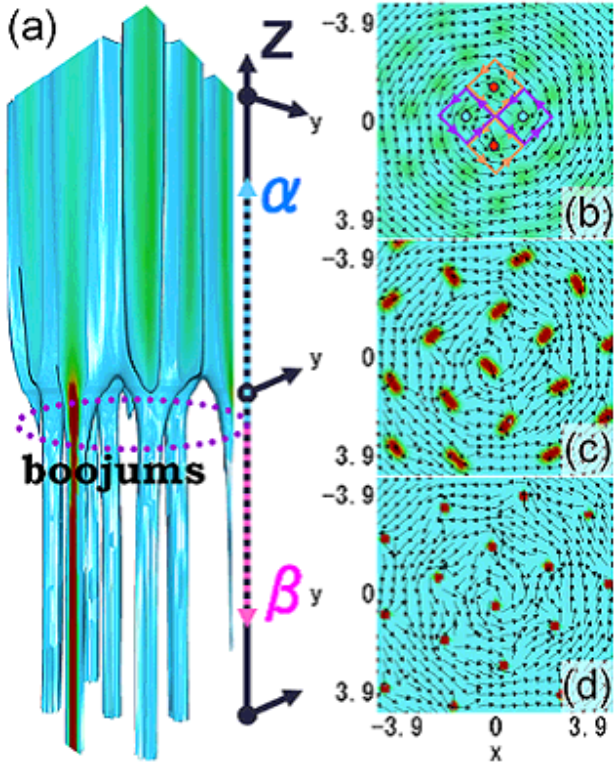}
  \caption{Boojums represented by vorticity of effective superflow.
 (a) Three-dimensional distribution of $|\Vec{\omega} _{\rm eff}|$ with several boojums near interface. (b-d) Distributions of $( \Vec{\omega} _{\rm eff}) _z$ and vectorial plot of unit vector $\Vec{v}_{\rm eff} / |\Vec{v}_{\rm eff}| $ in cross sections perpendicular to $\hat{\Vec{z}}$ at (b) $z=10.8$, (c) $z=-0.4$, and (d) $z=-11.2$. Red shows the region with high vorticity. Plot (b) shows the lattice of continuous vortices and four Wigner-Seitz cells in the $\alpha$-phase. Red and light blue points in the cells show the positions of the vortices of components 1 and 2, respectively. Plot (c) shows the pairing of continuous vortices near the interface. Plot (d) shows that singular vortices make a triangular lattice below the boojums ($\beta$-phase).}
\end{figure}

The bifurcation of vorticity through the boojum was found by investigating the circulation. The circulation per singular vortex in the $\beta$-phase is $2\pi$, just like that in a typical one-component BEC. The circulation around a continuous vortex was estimated by first assuming that the square lattice is periodic. Then, each Wigner-Seitz cell has a continuous vortex, and $n_1=n_2$ on the boundary of the cell \cite{n1=n2}. The circulation per continuous vortex is 
 \begin{eqnarray}
 \int_{\rm cell} \Vec{\omega}_{\rm eff} \cdot d\Vec{S}=
 \circ \hspace{-.93em}\int_{\rm boundary}
 \frac{n_1 \Vec{\nabla} \theta_1+n_2 \Vec{\nabla} \theta_2}{n_1 +n_2} \cdot d\Vec{l} \ \ \ \ && \nonumber \\
 = \frac{1}{2} \circ \hspace{-.99em}\int_{\rm boundary} (\Vec{\nabla} \theta_1 +\Vec{\nabla} \theta_2) \cdot d\Vec{l}
 =\pi . &&
\end{eqnarray}
 Thus, the whole circulation per singular vortex in the $\beta$-phase is transmitted to a pair of continuous vortices in the $\alpha$-phase. This process of vorticity bifurcation makes the velocity field in the $\alpha$-phase more similar to a solid-body rotation than that in the $\beta$-phase because the number of vortices of each component in the $\alpha$-phase is almost equal to that in the $\beta$-phase. This is observed by comparing the velocity fields in the $\alpha$-phase without singularities (Fig. 2(b)) and in the $\beta$-phase with singularities (Fig. 2(d)).

This consideration makes us think of the interesting structure of boojums of an $n$-component BEC; a singular vortex may split at a boojum ($n$-boojum) to $n$ continuous vortices with the same circulation $2\pi/n$.
 When $n \rightarrow \infty$, vorticity becomes uniform ($\Vec{v}_{\rm eff} \rightarrow \Vec{v}_{\rm solid}$, $\Vec{\omega}_{\rm eff} \rightarrow 2 \Vec{\Omega} $) above the lattice of $\infty$-boojums, as shown in Fig. 3(a). Such a topological structure of vorticity with $\infty$-boojums should generally appear on a boundary between one medium with singular vorticity and the other with a uniform vorticity. One example is the interface between a type-${\rm I\hspace{-.1em}I}$ superconductor and a normal media under a magnetic field perpendicular to the interface, where $\infty$-boojums of magnetic flux should appear.

It is useful to understand the topological structure of vorticity of a boojum to consider its analogy to a Dirac monopole. When we consider the vorticity $\Vec{\omega}_{\rm eff}$ as the magnetic flux density $\Vec{B}$, a Dirac monopole and a string respectively correspond to a boojum and a singular vortex \cite{volovik}. However, the radiation of $\Vec{\omega}_{\rm eff}$ from a boojum is different from that of $\Vec{B}$ from a Dirac monopole, as shown in Figs. 3(b) and 3(c).

\begin{figure}[th] \centering
  \includegraphics[width=.95 \linewidth]{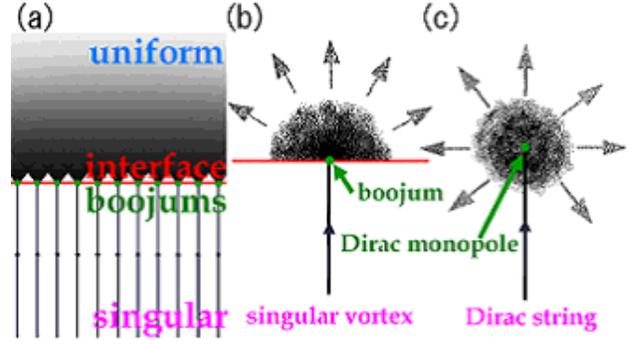}
  \caption{(a) Lattice of $\infty$-boojums. (b) Vorticity distribution around boojum. (c) Magnetic flux distribution around Dirac monopole.}
\end{figure}

  The pseudospin representation \cite{KTUeff} provides insights for understanding the similarity between two-component BEC and superfluid $^3$He.
 We consider a spin-$1/2$ BEC for the following spinor $\Vec{\Psi}=[\psi _1,\psi _2]^T=\psi_T \Vec{\chi}$ as the order parameter
 \begin{eqnarray}
 \psi_T &=&\sqrt[]{n_T} e^{i\theta_T /2}, \\
 \Vec{\chi} &=&\left[ 
 \begin{array}{c}
 {\rm cos}(\vartheta /2)e^{-i\varphi /2} \\
 {\rm sin}(\vartheta /2)e^{i\varphi /2} \\
 \end{array} 
 \right],
 \end{eqnarray}
 where $ n_T = n_1+ n_2$, $\theta _T =\theta _1+\theta _2$,
 ${\rm cos} \vartheta =(n_1-n_2)/(n_1+n_2)$, and $\varphi =\theta _2-\theta _1$.
 The parameters $\vartheta$ and $\varphi$ refer to the polar and azimuthal angles of the local pseudospin $\Vec{s}$ which is defined as
 $\Vec{s}=\Vec{\chi}^{\dagger}  \Vec{\sigma} \Vec{\chi}
 ={\rm cos}\vartheta \hat{\Vec{z}}+{\rm sin}\vartheta
  ({\rm cos}\varphi \hat{\Vec{x}}+{\rm sin}\varphi \hat{\Vec{y}})$, where $\Vec{\sigma}$ is the Pauli matrix and $\Vec{s}^2=1$.
 From these relations, the effective velocity $\Vec{v}_{\rm eff}$ of eq. (2) and its vorticity $\Vec{\omega}_{\rm eff}$ are 
\begin{eqnarray}
\Vec{v}_{\rm eff} =  \frac{1}{2i}
\frac{\Vec{\Psi}^{\dagger} \Vec{\nabla} \Vec{\Psi}
     -\Vec{\Psi} \Vec{\nabla} \Vec{\Psi}^{\dagger}}{\Vec{\Psi}^{\dagger} \Vec{\Psi}}
= \frac{1}{2} \Vec{\nabla} \theta _T - \frac{1}{2} {\rm cos} \vartheta \Vec{\nabla} \varphi, && \\
\Vec{\omega}_{\rm eff} 
=  \frac{1}{2} \Vec{\nabla} \times \Vec{\nabla} \theta _T \ \ \ \ \ \ \ \ \ \ \ \ \ \ \ \ \ \ \ \ \ \ \ \ \ \ \ \ \ \ \ \ \ \ \ \ \ \ \ \ \  && \nonumber \\
-\frac{1}{2}{\rm cos} \vartheta \Vec{\nabla} \times \Vec{\nabla} \varphi
+\frac{1}{2} (\Vec{\nabla} \vartheta) \times ({\rm sin} \vartheta \Vec{\nabla} \varphi). \ \ \ \ \ &&
\end{eqnarray}
 The first and second terms in eq. (7) vanish except at the vortex cores. The scalar product $[(\Vec{\nabla} \vartheta) \times ({\rm sin} \vartheta \Vec{\nabla} \varphi)] \cdot d \Vec{S}$ is equal to the infinitesimal solid angle $d \Vec{\Omega}$ covered by the $\Vec{s}$ orientations within the infinitesimal plane $d \Vec{S}$. As a result, the third term can be rewritten as
\begin{eqnarray}
\frac{1}{2} (\Vec{\nabla} \vartheta) \times ({\rm sin} \vartheta \Vec{\nabla} \varphi)
= \frac{1}{4}e_{ijk} s_i \Vec{\nabla} s_j \times \Vec{\nabla} s_k \equiv \Vec{q}(\Vec{r}),
\end{eqnarray}
 which is connected to the Mermin-Ho relation for the $\Vec{s}$ texture \cite{Mermin-Ho}. Figure 4 shows the pseudospin texture and the values of $\varphi$ in cross sections perpendicular to $\hat{\Vec{z}}$. In the $\alpha$-phase (Fig. 4(a)), the first and second terms cancel, leading to
 $\Vec{\omega}_{\rm eff} = \Vec{q}(\Vec{r})$. Thus, the circulation of eq.(3) requires that all points on the surface of a unit hemisphere, drawn with $\Vec{s}$ as the radius vector, are covered by the $\Vec{s}$ orientations within a Wigner-Seitz cell in Fig. 2(b), as observed in Fig. 4(d). The pseudospin texture in the $\alpha$-phase is similar to the texture of the order parameter $\hat{\Vec{l}}$ in the structure of a locked vortex (LV1) in a rotating superfluid $^3$He-A \cite{LV1texture}. When we approach the interface, spin domains are formed with a domain wall that connects two pairs of up-spin and down-spin (pairs of the original vortices), as shown in Figs. 4(b) and 4(e). A boojum appears at the point where the up-spin and down-spin annihilate. This annihilation removes domain walls, thus making the domains larger.
 In the $\beta$-phase, the spin domain walls vanish, which means that $\Vec{s}$ becomes uniform and lies in the $x$-$y$ plane; $\Vec{\chi}={\rm const}$. Thus, the $\beta$-phase is described only by $\psi_T$, which behaves similar to the order parameter of a one-component BEC such as superfluid $^3$He-B.

\begin{figure}[t] \centering
  \includegraphics[width=.95 \linewidth]{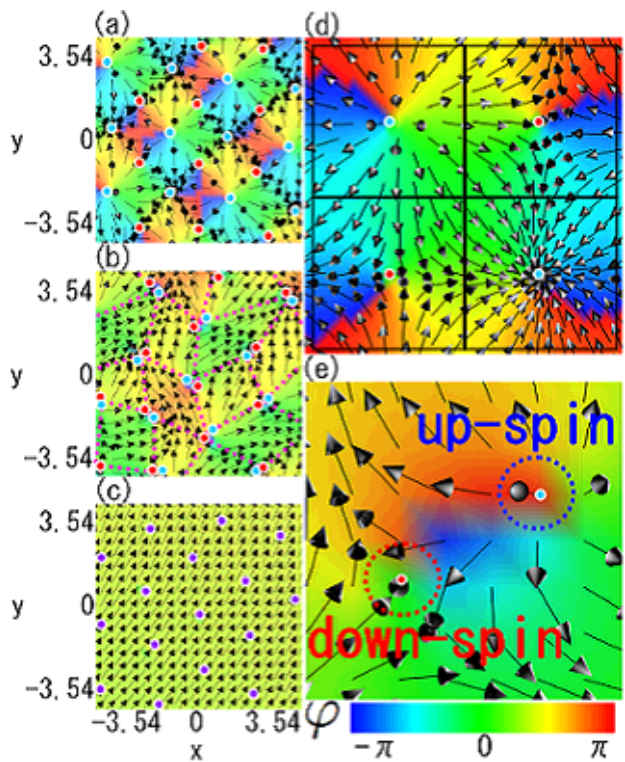}
 \caption{Textures of pseudospin $\Vec{s}$ and distribution of $\varphi$ (the azimuthal angles of $\Vec{s}$) in cross sections perpendicular to $\hat{\Vec{z}}$. (a) and (d) Cross sections at z=10.8. (b) and (e) Cross sections at $z=-0.4$. (c) Cross section at $z=-11.2$. Red and blue points show the positions of down-spin and up-spin at the vortex cores of components 1 and 2, respectively. (d) and (e) are magnified images of (a) (rotated by $45^\circ$) and (b), respectively.
 The domain walls are shown as pink dotted lines in (b). The four squares in (d) are the four Wigner-Seitz cells in Fig. 2(b).}
\end{figure}

In conclusion, we have numerically simulated boojums in two-component atomic BECs by making the intercomponent and intracomponent coupling constants spatially dependent. The structure of the boojums is analyzed by introducing effective velocity and pseudospin representation. In general, boojums should appear in rotating two-component BECs with the intercomponent interaction repulsive in one phase and attractive in the other. It is experimentally possible to control the interaction parameters by the Feshbach resonance tuned with a spatially dependent external field. Hence, we expect that boojums could be observed in two-component BECs. Studies on the dynamics of this system and BEC systems of more than two components are now in progress and will be reported elsewhere.

We acknowledge G. E. Volovik and R. H\"{a}nninen for useful discussions.
 MT acknowledges the support of research grants from the Japan Society for the Promotion of Science (Grant No. 15340122).


\begin{thebibliography}{99}
\bibitem{n1=n2} In Fig. 2(b), the straight line on every segment of the boundary of the Wigner-Seitz cell is a line of antisymmetry between $n_1$ and $n_2$ . Thus, $n_1=n_2$ on the boundary.
\end{thebibliography}
\end{document}